# Implicitly Modeling Frequency Control within Power Flow

Aayushya Agarwal, Amritanshu Pandey, Marko Jereminov, Larry Pileggi
Dept. of Electrical and Computer Engineering
Carnegie Mellon University
Pittsburgh, PA

*Abstract*—In this paper, we extend a circuit-based, current-voltage power flow formulation to include frequency deviations and implicitly model generator primary and secondary control actions as a function of their temporal dependence. This includes extending the slack bus generator model(s) to better represents its true behavior with frequency controls. These implicit models obviate the need for outer iteration loops and improve the robustness of the simulation convergence when frequency deviations are considered. The simulation framework is highly scalable and is demonstrated on 85k+ bus systems.

## I. INTRODUCTION

An electrical power grid represents a constantly changing interconnected network of synchronized generators and consumers. In order to effectively study and plan the operation of an interconnected grid, it is necessary to accurately simulate its steady-state response under various conditions, such as increased demand, contingency, etc. Realistically, when generation does not match the load demands, the grid is forced to change its frequency to stabilize the interconnected network [1]. This is a consequence of an inertial response of generators which provides necessary electric power by creating a difference in torque and as a result, decreases or increases the frequency of the grid. To resist this change, primary and secondary frequency control schemes adjust the real power of generators. These frequency control mechanisms are generally not modeled in traditional power flow, which is the steady-state analysis used in operation and planning of the bulk electric grid. In fact, power flow analysis [2] assumes that the primary frequency of the grid is always maintained at a nominal value (60 Hz or 50 Hz) and neglects the frequency dependencies of the grid models.

To satisfy the power balance between the generation, demand and line losses, the corresponding power flow problem generally incorporates one or more slack bus generators to provide the mismatch power that is needed to ensure simulation convergence. However, the slack buses do not reflect the actual operation of the grid, since in reality all generators (or a subset) respond to this power mismatch by changing their real power as a function of frequency. Therefore, to mimic the true behavior of the grid in power flow, it is not only necessary to include a frequency state variable while modeling the temporal behavior of primary and secondary control, but also to improve the model for the slack bus generator to mimic the true physical behavior of this generator.

To accurately characterize the state response of the network with frequency information, existing practices are generally based on running transient dynamic simulations by modeling the generator primary and secondary control loops [3], such as the COSMIC model in [4] and the quasi-dynamic model in [5]. However, transient analyses require a small time-step to ensure simulation convergence with accuracy. This results in long simulation times for steady-state solution, thereby making it unsuitable for operation of the grid or for bulk contingency analyses [6].

Due to its efficiency, power flow is often used for the majority of operation and planning studies of the grid. In governor load flow method [16], power flow has been extended to include frequency as a state variable to further improve the solution's accuracy. The authors in [9]-[16] further introduced new constraints and corresponding equations within the traditional "PVQ" formulation to represent the frequency deviation that adjusts the active power of generators through droop and Automatic Generation Control (AGC) mechanisms. Other advancements have also extended the current injection method to also include a frequency state and corresponding generator primary and secondary control mechanisms [12]. While these recent advancements of power flow formulations with frequency information have improved the accuracy of the steady-state solutions, they have not considered the temporal dependency between primary droop control and secondary AGC control.

To accurately obtain the final steady-state of the grid, it is important to consider the sequence in which the frequency control mechanisms occur. The authors in [9]-[14] did not incorporate primary and secondary control into the slack bus model, thereby the final response still relies on the slack power to overcome mismatches in the grid. In addition, the approaches in [9]-[16] have not demonstrated a formulation capable of scaling to large test cases, wherein it is difficult to distinguish between an infeasible grid case and one that is unable to converge due to lack of simulation robustness from ill-conditioning or complexity of the solution matrix [19]. One of the factors contributing to the lack of robustness is the use of outer loops to resolve violations of active power limits during primary control while using Newton-Raphson (N-R) based iterations. Outer loops have been used before to resolve discontinuous models such as PV-PQ switching, but were shown to result in oscillations and increased iterations [18].

In this paper, we incorporate frequency deviation information and frequency control mechanisms into a circuit theoretic current-voltage based power flow formulation [17]. The equivalent-circuit based power flow employs circuit simulation heuristics that enable robust convergence for large transmission and distribution systems, and is further able to identify infeasible grid cases [19] while locating and quantifying the weakest sections of the grid.

The primary contribution of this paper is the framework that extends the power flow formulation to robustly capture steady-state response of primary and secondary generator control through frequency state information. This approach implicitly models the generator control mechanisms (primary droop and secondary AGC) to include active power limits, thus removing the need for problematic outer loops [18]. We further incorporate realistic behavior in the power flow slack bus model to behave as the other generators in the system with regard to frequency deviation. The approach also captures the temporal sequence of events by considering multiple steady-states due to control actions. We demonstrate the efficacy of this framework on large scale transmission systems, namely the 85k+ nodes US Eastern Interconnection test cases.

## II. BACKGROUND

### A. Power flow in current and voltage state variables

The proposed framework in this paper builds on power flow formulations that model the grid in terms of current and voltage (I-V) state variables. Recent advances in power flow have shown that representing the transmission and distribution networks in terms of an equivalent circuit model [17] can be utilized to achieve robust convergence properties using circuit simulation techniques. Namely, the authors in [17] developed efficient step limiting heuristics and homotopy methods that enabled robust convergence of the network independent of the size or complexity. This equivalent circuit-formulism is also flexible to incorporate any other physics-based behavior such as frequency control.

### B. Frequency Dependent Loads

Two of the most prominent aggregated load models used within the power flow analysis are PQ and ZIP models and both are influenced by frequency. In [20] it was demonstrated that power flow load models can be extended to model the change in power consumed based on the frequency of the grid, by multiplying the total powers by a frequency term:

$$P_L = P_{PQ/ZIP}(1 + K_{pf}\Delta f) \quad (1)$$
$$Q_L = Q_{PQ/ZIP}(1 + K_{qf}\Delta f) \quad (2)$$

where $P_{PQ/ZIP}$ and $Q_{PQ/ZIP}$ are the active and reactive powers at the nominal frequency and are calculated as either a ZIP or exponential model. $K_{pf}$ and $K_{qf}$ describe the linear relationship between the load parameters and a change in frequency, $\Delta f$.

## III. FREQUENCY CONTROL MODELS

Frequency deviations in the grid are generally contributed to a synchronous generator inertial response that supplies excess power required to satisfy power mismatch in the system. To prevent the decline of frequency due to this inertial response, primary (droop) control actions are first activated. Next, in order to fully restore the frequency towards the nominal one, secondary (AGC) control actions are then subsequently applied by a set of participating generators. The steady-state behavior of the primary as well as the secondary control actions can be macro-modeled as a change in generator's active power, $\Delta P$. Importantly, both the primary and secondary control can be defined in terms of variables corresponding to a change in active power due to their control mechanisms, $\Delta P^p$ and $\Delta P^s$ respectively. To incorporate these changes in active power produced into the power flow equations, we add an extra equation for each generator:

$$P_G = P_G^{SET} + \Delta P_G^p + \Delta P_G^s \quad (3)$$

While the aforementioned frequency controls have been previously incorporated into power flow [9], in the following sections, we introduce implicit models for the same and model the time dependence between primary and secondary controls.

### A. Primary Frequency Control

The active power change due to primary frequency control is linearly related to frequency [1]-[7] by:

$$\Delta P_G^p = -\frac{P_R}{R}\Delta f \quad (4)$$

where $P_R$ and $R$ are parameters describing the inertial and droop response of a generator. These parameters are dependent on the type/size of the generator and implementation of the droop control feedback [1].

To facilitate the use of primary control models, it is important to consider a generator's active power limits, given by:

$$\Delta P_G^{MIN} \leq \Delta P_G^p \leq \Delta P_G^{MAX} \quad (5)$$
$$\Delta P_G^{MIN} = P_G^{MIN} - P_G^{SET} \quad (6)$$
$$\Delta P_G^{MAX} = P_G^{MAX} - P_G^{SET} \quad (7)$$

where $P_G^{MAX}$ and $P_G^{MIN}$ are a generator's active power limits and $P_G^{SET}$ is the set active power specified by the test case.

It can be shown that the existing methods that include frequency deviation variable within power flow [16] handle the active power limits using discontinuous models that are resolved in an outer loop. However, discontinuous models are known to suffer from convergence challenges while running NR. These models are shown to oscillate or diverge [18] and can pose significant difficulties for convergence of large complex or ill-conditioned test systems. Therefore, to avoid these issues due to the use of a discontinuous model, a continuous implicit model given in (8) is developed to limit the real power of the generators while applying the primary control relationship in (4). We have previously incorporated implicit functions to model operational limits [18], such as for reactive power of the generators. These models help achieve robust convergence by not using outer loops, which have been shown to result in oscillations and increased iterations.

$$\Delta P_p = \begin{cases} \Delta P_{MIN}, & \text{Region 1} \\ a_{min}\Delta f^2 + b_{min}\Delta f + c_{min}, & \text{Region 2} \\ -\frac{P_R}{R}\Delta f, & \text{Region 3} \\ a_{max}\Delta f^2 + b_{max}\Delta f + c_{max}, & \text{Region 4} \\ \Delta P_{MAX}, & \text{Region 5} \end{cases} \quad (8)$$

To preserve the convergence properties of NR method, it is required that the functions that are solved for are first-order continuous. Equation (8) is a piecewise continuous function with a continuous first derivative, that enables it to apply NR to the problem without the need for challenging outer loops. The piecewise equation is segmented into five regions shown in Figure 1, each corresponding to a different operating state. Region 3 of the equation is the linear region which models the frequency dependence of (4) when the active power is within the operating limits. Regions 1 and 5 bound the generator's active power to the respective operating limits. In order to ensure first derivative continuity, Regions 2 and 4 are quadratic regions that patch the discontinuity of Region 3 and the bounds. The coefficients for the quadratic functions are found during initialization of the problem by solving for values that will match the first derivative and the function value at the points of intersection. This translates to a set of equations with 3 variables ($a_{min/max}, b_{min/max}, c_{min/max}$) for each quadratic region, as well as the frequencies at which the quadratic region intersects with the adjacent region.

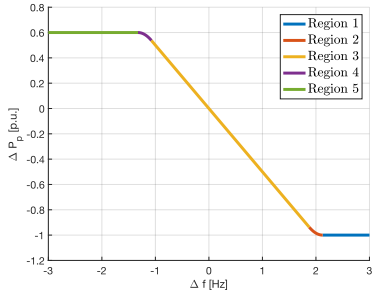

Figure 1: Implicit model for primary control.

### B. Secondary Frequency Control

The primary frequency control arrests the decline of frequency; however, it is not able to restore the frequency to nominal. Therefore, secondary frequency control (AGC) is applied to bring the frequency closer to nominal. AGC is governed by the area control error (ACE) that signals the generator to produce more or less active power, and is defined by:

$$ACE = (\Sigma P_{scheduled} - \Sigma P_{actual}) + 10\beta \Delta f_1 \quad (9)$$

where $\beta\ [\frac{MW}{0.1Hz}]$ is the frequency bias constant, and $\Delta f_1$ is the frequency deviation from nominal measured when the secondary control is activated [9]. Also, $(\Sigma P_{scheduled} - \Sigma P_{actual})$ measures the deviation of the net exchange of active power between areas from the scheduled net exchange. All the generators participating in AGC are collectively trying to minimize ACE, with each generator contributing a portion of active power governed by a participation factor, $\kappa$, as given in (10). An important note is that the ACE is calculated for the network before the AGC is enacted.

$$\Delta P_G^s = \kappa * ACE \quad (10)$$

### C. Modeling Frequency Controls in Slack Bus

To further model the frequency control mechanism for slack generators, the same implicit models of (8) for primary and secondary control (9) are added to each slack bus in power flow. An additional constraint for the slack generator's current ($I_R^s$ and $I_I^s$) and voltage ($V_R^s$ and $V_I^s$) state variables is included with corresponding variable as delta frequency ($\Delta f$):

$$(P_S^{SET} + \Delta P_S^p + \Delta P_S^s) = V_R^s I_R^s + V_I^s I_I^s \quad (11)$$

This equation improves the model of a slack generator by constraining it to produce a set amount of active power which can be influenced by the frequency of the grid. As a result, the slack bus is not the only generator liable for supplying the mismatch power in the system. Rather, we are able to model the slack bus as a PV bus by setting its output power to share the "missing" power in the system by all generators using frequency variable as the shared variable by (11) and set the voltage.

### D. Temporal Dependency of Frequency Controls

While at steady-state the primary and secondary control actions are both modeled to stabilize the frequency of the grid, in reality they have temporal properties. Figure 2 illustrates the frequency recovery of a grid experiencing a disturbance (at $t_1$), which then activates the primary control (resulting in the steady-state at $t_2$) and the secondary control (resulting in the steady-state at $t_3$).

The primary controller typically has a small time constant and reacts within seconds [7], however does not fully restore the frequency of the grid to nominal, as shown at time $t_2$ in Figure 2. Therefore, a secondary control mechanism, controlled by the ACE computed at $t_2$, is activated. This secondary controller has a much larger time constant and reacts within minutes after the frequency decline [7]. After the AGC has adjusted the active powers of the participating generators based on the ACE at $t_2$, a new steady-state is found at $t_3$, which minimizes the total ACE of the system. In case there is a network disturbance or a topology change (as is the case before every new dispatch), the whole simulation is performed again considering primary and secondary control in a sequential manner, as shown by the network disturbance at $t_3$. By considering multiple steady-states and using the previous solution as an initial condition for the next power flow, we are able to approximate the time dependence of the frequency controls.

On the other hand, existing methods typically solve for the steady-state of both primary and secondary controls together without considering the time dependence. This approach ignores the ACE value at $t_2$ and any network changes that may occur.

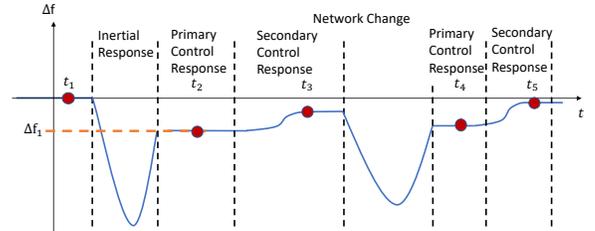

Figure 2: Time sequence of frequency due to frequency controls.

## IV. FREQUENCY DEPENDENT POWER FLOW FRAMEWORK

To find the steady-state of the grid while considering the primary and secondary frequency controls, we incorporate the aforementioned frequency control models into our power flow solver. We develop a framework around the models that mimics the dynamic behavior of the system by solving for the steady-states after each control action has occurred while respecting the sequence of events as illustrated in Figure 2. The algorithm, shown in Figure 3, activates each time a new file is obtained from the network topology estimator. For instance, in PJM ISO a new network file is obtained few minutes [22].

In the stage I of the simulation framework, the simulator applies the primary control response to arrest any decline in the system frequency due to the power mismatch in the system using the implicit primary control model for the generators (shown in Section III.C.1). The solution of this problem corresponds to the state at time $t_2$ in Figure 2.

In the stage II of the simulation framework, the simulator applies the primary control along with the secondary control to further bring frequency closer to the nominal. The input to the secondary control (i.e. ACE) is computed based on the frequency information from first stage and the models for the secondary control of generators include those that were developed in Section III.C.2 of the paper. The solution of this problem corresponds to the state at time $t_3$ in Figure 2.

Upon receiving a new network topology file, the simulator re-runs the Stage I and Stage II. By considering the temporal sequence of primary as well as secondary control the proposed approach better mimics the reality.

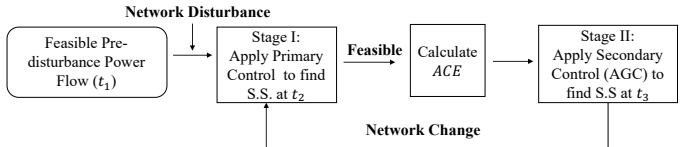

*Figure 3: Flow Chart for frequency-based Power Flow algorithm.*

## V. SIMULATION RESULTS

In this section, we demonstrate the use of frequency control models on a 23 bus testcase, savnw [1] when simulated under a contingency. The inertia and participation factor (for AGC) parameters of each generator are shown in TABLE 1. In addition, we modeled the frequency dependence of loads by assigning $K_{pf}$ as -80 MW/Hz and $K_{qf}$ as -220 MW/Hz.

TABLE 1 SAVNW TESTCASE GENERATOR PARAMETERS

| Generator ID | $P_{max}$ [MW] | $P_{min}$ [MW] | $\frac{P_R}{R}$ [MW/Hz] | $\kappa$ | Area | AGC |
|---|---|---|---|---|---|---|
| 101 | 890 | 0 | 950 | 0.5 | 1 | ✓ |
| 102 | 890 | 0 | 950 | 0.5 | 1 | ✓ |
| 206 | 990 | 0 | 660 | 0.5 | 2 | ✓ |
| 211 | 620 | 0 | 500 | 0.5 | 2 | ✓ |
| 3011 | 128 | 0 | 1500 | 0.8 | 5 | ✓ |
| 3018 | 1000 | 0 | 500 | 0.2 | 5 | X |

### A. Advantages of Implicit Model

In this result we utilize the 23-bus model described above and demonstrate that implicit models can achieve convergence for cases that otherwise fail with the use of discontinuous models in the outer loop. In the following scenario, the savnw testcase was modified by increasing the load by 20%, while increasing the $\frac{P_r}{R}$ for generator 101 to 1500 MW/Hz, which forced the generator to produce more active power. When simulated with the use of implicit models given in Section III.C.1, the system converged within 40 iterations from flat-start and generator 101's real power saturated at its maximum value. However, when the same case was solved using discontinuous models that resolved violations with an outer loop, oscillations were observed as shown in Figure 3. The oscillations were a result of two generators (101 ,102) violating their active power limits in the inner loop, and the simulation did not converge within the maximum iteration count.

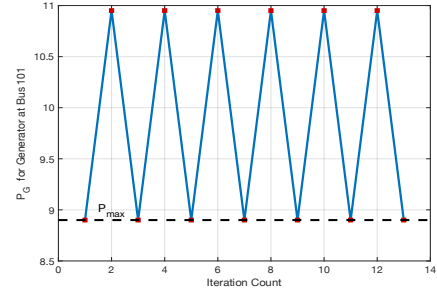

*Figure 4: Oscillations in savnw testcase with discontinuous model.*

### B. Primary and Secondary Controls on Savnw

In the following scenario, we apply a disturbance to the savnw test case while implicitly modeling the frequency controls. The simulation begins by solving for the state at $t_1$ in Figure 2, which is the case prior to the contingency (the active powers for the generators are listed in Table 2). The generator on bus 211 is then disconnected, causing an imbalance of power, and activates the primary control. The steady-state due to this control at $t_2$ is solved and the generators' real power is reported in TABLE 2 along with the frequency deviation. At this point we compute total ACE for the system, which is used as a measure for the secondary control. To find the steady-state due to both primary and secondary control at t=$t_3$ , we run another instance of power flow with the implicit primary and secondary AGC controls applied. The steady-state solution further restores frequency closer to nominal (Δf=-0.009 Hz), as shown in last column of TABLE 2.

TABLE 2: REAL POWER DURING PRIMARY AND SECONDARY CONTROL.

| | Real Power Generation $P_G^{SET}$ [MW] | | |
|---|---|---|---|
| Generator ID | **Pre-contingency** | Post-contingency | |
| | ($t_1$) | **Primary Control ($t_2$)** | **Secondary Control ($t_3$)** |
| 101 | 770 | 885 | 870 |
| 102 | 770 | 886 | 870 |
| 206 | 820 | 900 | 930 |
| 211* | 500 | - | - |
| 3011 | 240 | 421 | 472 |
| 3018 | 60 | 118 | 110[#] |
| Δf [Hz] | 0 | **-0.12** | **-0.009** |
| ACE (MW) | 0 | **12.7** | **0.93** |

### C. Simulating a Dynamic Power Network

While simulating the effect of different controls on the grid, it is also important to consider a change in network topology that will cause a dynamic response. Generally, a network file is generated every few minutes describing the changes in the system [22].

In the following scenario, after the steady-state at $t_3$ in the previous scenario, we simulate a change in the network by increasing all the loads by 5%. The change in network initiates the primary control (at $t_4$) and then the secondary control to reach new steady-states at $t_5$, as shown in Figure 2. The effect on the primary frequency of the entire simulation is shown in Figure 2 and ACE are shown in Table 3, which highlights the capability of the framework to respond to changes in the network and reapply the frequency controls, thereby simulating the quasi-transient nature of the power grid using power flow.

TABLE 3: FREQUENCY AND ACE VALUES AT VARIOUS TIME POINTS.

|  | $t_1$ | $t_2$ | $t_3$ | $t_4$ | $t_5$ |
|---|---|---|---|---|---|
| Δf | 0 | -0.12 | -0.009 | -0.087 | -0.006 |
| ACE | 0 | 12.7 | 0.93 | 9.13 | 0.65 |

### D. Scalability

Using the circuit heuristics, we are able to solve large and complex power systems robustly [17]. We further extend these properties to frequency dependent proposed formulation as shown in Figure 3. To demonstrate robustness, each testcase was modified by disconnecting a generator and was shown to converge starting from an arbitrary initial condition while modeling the primary and secondary primary controls. The cases were run on a single core on a 2.6GHz Intel i7 processor. The largest testcase shown in Figure 5 represents the Eastern Interconnect with 85k+ buses.

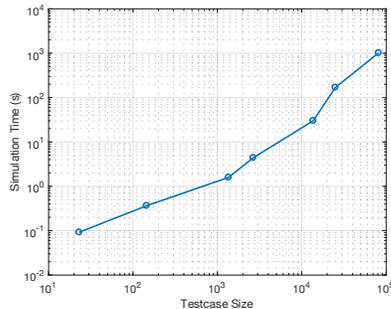

*Figure 5 Scalability of frequency dependent Power Flow.*

## VI. CONCLUSIONS

In this paper we introduced implicit frequency dependent power flow models that can be utilized to solve for the steady-state of the primary control response and the secondary frequency regulation. We extended the power flow circuit formalism to simulate realistic steady-state behavior by taking the system frequency of the grid into account. The steady-state responses were shown to provide realistic behavior for contingency analyses and were scalable to large systems. By accounting for the sequence of control actions that take place to reach a final steady-state, we more closely match the steady-state response to that from transient analysis simulation, thereby helping to unify these otherwise disparate formulations.

## VII. ACKNOWLEDGMENT

This work was supported in part by the Defense Advanced Research Projects Agency (DARPA) under award no. FA8750-17-1-0059 for RADICS program, and the National Science Foundation (NSF) under contract no. ECCS-1800812.